\documentclass[twocolumn,showpacs,preprintnumbers,amsmath,amssymb]{revtex4}

\usepackage{graphicx}
\usepackage{dcolumn}
\usepackage{bm}

\begin{document}

\title{Noise cancellation effect in quantum systems}
\author{Paolo Solinas,$^{1}$ Nino Zangh\`{\i}$^{1}$}

\affiliation{
$^1$ Istituto Nazionale di Fisica Nucleare (INFN) and
Dipartimento di Fisica, Universit\`a di Genova,
Via Dodecaneso 33, 16146 Genova, Italy}

\date{\today}

\begin{abstract}

We  consider the time evolution of simple  quantum systems under the
influence of random fluctuations of the control parameters.  We show
that when the parameters fluctuate sufficiently fast, there is
a cancellation effect of the noise. We propose that such an effect
could be experimentally observed  by performing a simple experiment
with
trapped ions. As a byproduct of our analysis,
we provide an explanation of the robustness against random perturbations
of adiabatic population transfer techniques in atom optics.
\end{abstract}


\maketitle
\section{Introduction}
In the last ten years the interest around the control
and manipulation of quantum  systems has grown very fast. The
possibility to encode and
process information has lead to innovative proposals.
The major results have been achieved in Quantum Information Processing
(QIP),
including both theoretical and experimental ones.
Quantum cryptography \cite{crypto} and
information transfer protocols \cite{teleportation} have enhanced our
understanding of information processing and this in the near future will 
presumably lead to a significant technical advancement.
Quantum Computation (QC) is still in a initial stage:
even if the quantum computers seem to be able to solve quickly some
problems which are intractable with classical computers \cite{algos},
more quantum algorithms are required to extend its applicability.

Unfortunately, the quantum system are very delicate and they are
subject to two
different kinds of errors. On one hand, there is the loss of
information due to decoherence the unavoidable interaction of quantum systems 
with their environments.
This problem has been extensively studied over the past few years
and  proposals to overcome it  have been put forward (and a few have
been experimentally tested).
These proposals include error avoiding \cite{error_avoiding},
error correcting strategies \cite{error_correcting} and
decoupling techniques \cite{bang-bang}.
The other source of errors is the imprecise control of the parameters
which perform the quantum operation (e.g. the laser or the magnetic field).
How to handle such errors is an open problem, though some progress has
been made  in the framework of the so called geometrical
quantum computation \cite{geometric_QC, non_adiabatic, HQC_noise}. 
The goal of this paper is to approach the second problem in very simple and 
idealized situations.

A simple way to model the parameter noise is to consider a quantum
system subject to a stochastic fluctuating field with zero mean.  Such a model
has been recently  considered in order to study the effect of the
noise on  holonomic quantum gates: in Ref. \cite{par_noise}  it has been
shown that there is a cancellation effect  for a fast fluctuating stochastic
field and shown that such a cancellation  is due to the geometrical
dependence of the holonomic operator.
Recently, the general validity of such cancellation effect has been clarified:
in Ref. \cite{facchi} is has be shown that for sufficient fast fluctuating
stochastic field with zero mean the effects of the noise are wiped out.

In this paper we shall study simple quantum systems
subjected to stochastic noise and discuss some applications.
In Section \ref{sec:2ls}  we consider random perturbations
which are  diagonal in the logical basis and
propose an experiment to test the cancellation effect of the the
noise|a simple modification of the experiment done
by Kielpinski {\it et al.} \cite{kielpinski}.
Then we consider a more general noise and compute the
{\it fidelity}. By elementary perturbation theory, we show that the
effects of the noise are wiped out.
In Section \ref{sec:adiabatic} we discuss how noise cancellation
could be relevant for {\it adiabatic population transfer} experiments in 
atoms optics (in this case, the cancellation effect avoids the break of the 
adiabatic approximation, and allows for desired transformation, even in
presence of fast fluctuating noise).

For sake of simplicity, all the simulations and analytical calculation
are done with Gaussian noise distributions but presumably
analogous results can be achieved with a generic
stochastic noise with zero mean \cite{facchi}.

\section{Two-level Systems}
\label{sec:2ls}

Consider a two-level system evolving according to Hamiltonian
\begin{equation}
H(t)= H_{0} + \delta H_{I}(t)
\label{eq:diagonal_noise}.
\end{equation}
   Suppose  that
$H_0 = B_z{\sigma}_z$ (with $\sigma_{z}$ being a Pauli matrix) and that
$\delta H_{I}(t)$  is of the form
\begin{equation} \delta H_{I}(t) =  \sum_{j=0}^N \delta A^j S_{j}(t)\,,
\label{eq:hit}
\end{equation}
where $\delta A^{j}$ are random $2\times2$ matrices  and $S_{j}(t)$ are
``box functions'' with time step $\tau$, i.e., they are functions
equal to 1 in the time interval $(j\tau, (j+1)\tau)$
and zero otherwise. Let $T$ be some ``final'' time at which we wish to
consider the system,  and  let $\tau = T/N$;
$\tau$ can be regarded  the correlation time of the noise. Then the
evolution operator from time zero to time $T$, generated by $H(t)$ with
$\delta H_{I}(t)$ given by (\ref{eq:hit}), can be written as
\begin{equation}
U(T) = U_{N}U_{N-1}\cdots U_{j}\cdots U_{2}U_{1}
\label{cron}
\end{equation}
where
\begin{equation}
U_j =  \exp\left[-i\tau\left(B_z{\sigma}_z+ \delta A^j
\right)\right]\,.
\label{cronj}
\end{equation}

\subsection{Diagonal Noise}
\label{sec:diagonal_noise}
First, we shall focus on the very simple case
corresponding to the choice $\delta A^{j}= \delta B^j_z \sigma_{z}$,
where the $\delta B^j_z$, $j=1,\dots,N$ are independent Gaussian distributed
random variables with mean zero. In this case the noise is diagonal in
the logical basis $\{|\uparrow\rangle , |\downarrow\rangle\}$ in which
$\sigma_{z}$ is diagonal. Then (\ref{cron}) is trivially computed (due
to the commutativity of the $\delta A^{j}$),
\begin{equation} U(T) = U^{0}(T)  \delta U (T)\,
\label{ut}
\end{equation}
where $U^{0}(T)= \exp(-i TH_{0})$ is the evolution generated by
$H_{0}$, and
\begin{equation}
    \delta U(T) = \left ( \begin{array}{ccc}
               e^{-i \sum_j \frac{\delta B^j_z}{N} T}       &  0\\
               0       &  e^{i \sum_j \frac{\delta B^j_z}{N} T}
               \end{array} \right )
               \label{dut}
\end{equation}

A standard  performance estimator is the {\it fidelity} \cite{nielsen},
which, in our case, is given by
\begin{equation}
    \mathcal{F}_T = \sqrt{\left|\langle \psi^{(0)}(T) 
      |\psi(T)\rangle\right|^{2} }
    \label{eq:fidelity_def}
\end{equation}
where $\psi^{(0)} (T)$ is the ``ideal'' final  state evolved according
to $H_{0}$ (that is, when the noise is turned off) and $\psi (T)$ is
actual final state (that is, when the noise is turned on). The two
final states are generated, of course, by  the {\it same} (generic)
initial state  $|\psi(0)\rangle = \alpha |\uparrow\rangle
+ \beta |\downarrow\rangle$.  From (\ref{ut}) and (\ref{dut}) one easily  
obtains
\begin{equation}
    \mathcal{F}_T^2(\chi, \alpha, \beta) = |\alpha|^4 + |\beta|^4 + 2
|\alpha|^2|\beta|^2 \cos\chi \,,
\label{eq:fidelity_general}
\end{equation}
where \begin{equation}\chi = 2 T \frac{1}{N}\sum_{j=0}^N \delta
B^j_z\,.  \label{eq:chi}
\end{equation}

\begin{figure}[t]
    \begin{center}
      \includegraphics[height=5cm]{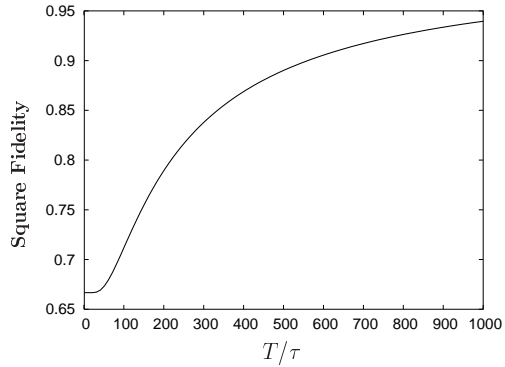}
      \caption{\label{fig:2ls_diagonal} The {\it square fidelity} from  
	noisy process described by (\ref{eq:diagonal_noise}).
        The solid line is the theoretical mean square fidelity, eq.
        (\ref{eq:fidelity}), for $T=100$ (in arbitrary units) and
        ${\delta B}=0.1$.}
    \end{center}
\end{figure}

Since the random variables $\delta B^j_z$ in (\ref{eq:chi}) are taken
to be independent and identically  distributed according to
a Gaussian with zero mean and variance ${\delta B}^2$,  the probability
distribution of $\chi$ is
\begin{equation}
    \mathcal{P}(\chi) = \frac{\sqrt{N}}{\sqrt{2\pi} 2 T {\delta B}}
e^{- \frac{N \chi^2}{2 (2T)^2 {\delta B}^2}}\,,
\label{eq:mean_over_realization}
\end{equation}
By averaging (\ref{eq:fidelity_general}) with respect to
(\ref{eq:mean_over_realization}),
we arrive at the mean square fidelity:
\begin{eqnarray}
    \mathcal{F}_T^2(\alpha, \beta) &=& \int d \chi \mathcal{P}(\chi)  
\mathcal{F}_T^2(\chi, \alpha, \beta)= \nonumber \\
    &=&  |\alpha|^4 + |\beta|^4 + 2 |\alpha|^2|\beta|^2 e^{-\frac{2 T^2  
{\delta B}^2}{N}}
    \label{eq:fidelity}
\end{eqnarray}

As it can be easily seen, for $N \rightarrow \infty$ and $\tau\to 0$
(while keeping $N\tau =T$ of order 1),  the mean square fidelity
$\mathcal{F}_T^2(\alpha, \beta)$ approaches $1$
(since $|\alpha|^{2}+|\beta|^{2}=1$), independently of the initial  
state.
This corresponds to  a short correlation time of the noise, that is, to a 
fast random fluctuating field.

Now we'd like to provide some perspective on  such limit behavior.
Of course,
the limit $\tau \rightarrow 0$ is  only an idealization for
$\tau$ small but finite.
The above results are obtained for {\it constant} $\delta B^2$,  but
the Heisenberg uncertainty relation imposes strong constraint on the
energy fluctuation happening in such a short time interval,  
$\delta B^2 \propto 1/\tau$. Note that for $\delta B^2 \propto N$
eq. (\ref{eq:mean_over_realization}) and (\ref{eq:fidelity}) describe
a system interacting with a white noise environment
that is the standard way to model the decoherence effect.
So, for very small $\tau$   the variance $\delta B^2$  should not be
any more considered constant
and  our approximation of constant variance breaks down.
However,  rather than microscopic (environmental) noise, we are
interested in modeling the macroscopic (parametric) noise due to imprecision 
in the control field, and  for such our approximation should apply.
In this case it is  interesting  to ask  whether  the condition of small $\tau$
could  be physically relevant.
In general, in a quantum evolution the final time $T$ is fixed; the
correlation time $\tau$ can be hardly controlled (i.e., stabilizing the
control field) and we are not in the condition to have a cancellation effect.
In Section \ref{sec:experiment} we present an experimental proposal to
test the presence of this effect; the experiment realizability lies in the
control of the correlation time $\tau$ of the {\it simulated} noise.
Moreover, there are situations in which this effect can be
experimentally relevant: when we have a further degree of freedom and can 
change the evolution time $T$.
In these cases, fixed $\tau$ with the above properties, we can prolong
$T$ in order to have $T \gg \tau$ (that is $N \gg 1$) and exploit the
cancellation effect.
We give an example of such situation in a adiabatic evolution in Section
\ref{sec:adiabatic}.

We observe that another interesting  feature of eq.  
(\ref{eq:fidelity})
is that the cancellation of the noise does not depend on the strength
$\delta B$: it is always possible to find a suitable $N$ in order to
obtain cancellation of the noise|invert equation (\ref{eq:fidelity})
and express $N$ in terms of ${\delta B}$ for given  $\mathcal{F}_T$
and evolution time $T$. 

We recall that dependence of the  
fidelity on a specific choice of the initial state is
usually eliminated by  averaging
$\mathcal{F}_T^2 (\alpha, \beta)$ over all the possible initial states  
with respect to the
uniform distribution on the unit sphere in the Hilbert space of the  
system \cite{nielsen} in our case the (projective) sphere  
$|\alpha|^{2}+|\beta|^{2}=1$. Performing this operation yields to
  \begin{equation}
  \mathcal{F}_T^2 \equiv <\mathcal{F}_T^2 (\alpha, \beta) > =  
\frac{1}{3}(2 +  e^{-\frac{2 T^2 {\delta B}^2}{N}})
    \label{eq:mean_over_state}
  \end{equation}

\begin{figure}[t]
    \begin{center}
      \includegraphics[height=5cm]{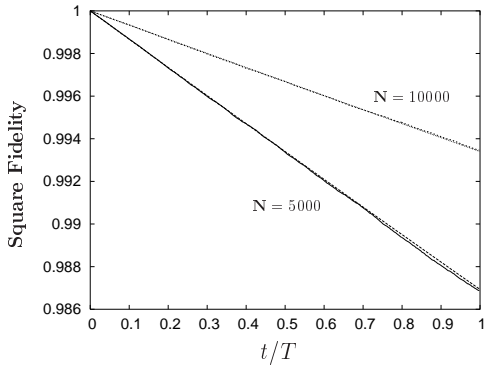}
      \caption{\label{fig:deco_time} The decoherence effect during the
        evolution for two different value of $N= T/\tau$.
        Theoretical curves (dashed line) and those obtained by numerical 
	simulations (solid line) are showed. For the theoretical curve 
	(eq. \ref{eq:deco_expression}) the decoherence time is
        $t_{deco} = \frac{N}{2T^2 {\delta B}^2}$.
      }
    \end{center}
\end{figure}

The simple model  we have been considering here is often used as a toy  
model for phenomenological decoherence \cite{phen_deco, scully}.
The relationship between noise cancellation  and
decoherence is  easily seen by considering the time evolution at the
discrete times $t=k \tau$, $k \le N$. By proceeding as before,
the average square fidelity at time $t=k \tau$ results
\begin{equation}
   \mathcal{F}^2_t =  \frac{1}{3}(2 + e^{-\frac{2 T t {\delta B}^2}{N}})
\label{eq:deco_expression}
\end{equation}
with which it is quite natural to associate a
`decoherence time' $t_{deco}=  \frac{N}{2T {\delta B}^2}$
(see Figure \ref{fig:deco_time} \cite{sample,note_on_realizations}).

\subsection{Experimental proposal}
\label{sec:experiment}

In Ref. \cite{kielpinski} Kielpinski {\it et al.} used the same idea
to simulate the decoherence effect due to interaction of the quantum
system with the environment degree of freedom.
The authors used trapped ions and study the
coherence of superposition of quantum state subject to simulated noise.
The logical states were the hyperfine states of a trapped Beryllium ion
$|F=2, m_F=-2\rangle$ and $|F=1, m_F=-1\rangle$ sublevels of the ground
state $^2$S$_{1/2}$. The environment noise was simulated shining the  
ions
with a off-resonant laser with random varying amplitude for the
electromagnetic field $E^j$ and random intensity (proportional to  
$(E^j)^2$).
The laser electromagnetic field produce a AC Stark effect on the ions
and let one state to acquire a random phase respect to the other.
This effect is {\it quadratic} ({\it quadratic Stark effect})
in the electromagnetic field $E^j$.
In fact, the two hyperfine states have the same angular momentum
(they have both $L=0$) and the difference is in the spin part of the wave 
function.
The splitting of the energy level is {\it linear} ({\it linear Stark
effect})
for state with different angular momentum since only these states have
non-vanishing matrix element
$\langle L=i| \vec{E} \cdot \vec{r} |L=k \rangle$ (with $i\neq k$).
Then in the above example we have corrections to the energy level only
quadratic and not linear in $E^j$.
Because of the quadratic energy shift, we have a random phase difference
proportional to $(E^j)^2$ for every $\tau$ interval.
Even if this effect is sufficient to produce decoherence effect
(as found by the authors), we cannot presumably see the cancellation effect 
discussed above in a transparent way since our new stochastic variable 
$\chi \propto \sum (E^j)^2$ has no zero mean.

A small modification of this experiment should allow us to see sharply 
this cancellation effect.
To have an evolution described by Hamiltonian (\ref{eq:diagonal_noise})
it is sufficient to use states with different angular momentum in order to 
produce a {\it linear Stark effect}. At every time interval $\tau$ the
perturbation of the laser produce a shift of the energy levels proportional 
to the random intensity of the laser $E^j$; this produces an
evolution where the phase difference between the states is given
by a known dynamical part plus a random phase $exp(2i \alpha E^j \tau)$
(where $\alpha$ is a proportional constant).
In this case, the new stochastic variable $\chi \propto \sum_j E^j$
has still  zero mean and we expect to obtain results shown in the
previous section: fixed the evolution time $T$ we should see an increase of
$\mathcal{F}^2$ as the correlation time decreases (see Figure
\ref{fig:2ls_diagonal}).
Moreover, if the environment decoherence does not depend on the  
simulated noise, we should be able to see effect analogous to the ones in 
Figure \ref{fig:deco_time}.
In particular, subtracted the effect of the environment, the decoherence
time should increase as $t_{deco} \propto \frac{1}{\tau}$.

\subsection{Off-Diagonal Noise}
\label{sec:non_diagonal}

\begin{figure}[t]
    \begin{center}
      \includegraphics[height=5cm]{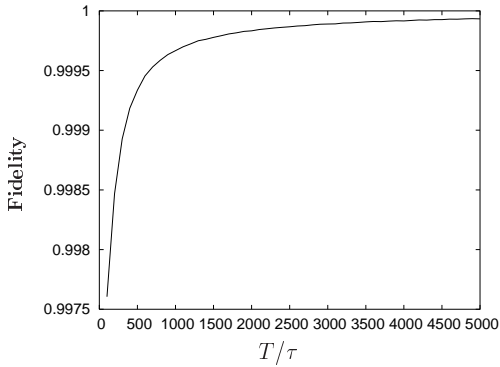}
      \caption{\label{fig:non_diagonal} Cancellation effect
        for system subject to off-diagonal noise 
	discussed in section \ref{sec:non_diagonal}.
        For the stochastic process $\delta B=0.01$. 
	When the number of random extraction  
	increases the {\it fidelity} approaches $1$.
      }
    \end{center}
\end{figure}

We now consider the case of off-diagonal noise, that is,
the matrices $\delta A^{j}$ in (\ref{eq:hit}) are of the form
$\delta A^j= \delta B_x^j \sigma_x + \delta B_y^j \sigma_y$,
with $\delta B^j_x$, $\delta B^j_y$, $j=1,\dots,N$  independent
Gaussian distributed random variables with mean zero, and $\sigma_x$,
$\sigma_y$ being the usual Pauli matrices. Then the one-step evolution
operators $U_j$ in (\ref{cron}) are
$$U_j = e^{-i \vec{B}^j \cdot \vec{\sigma} \tau} =
\cos(B^j\tau) - i \hat{n}^j \cdot \vec{\sigma} \sin(B^j\tau)\,,$$
where $B^j$ is the modulus of the vector $\vec{B}^j= (\delta B_x^j,
\delta B_y^j, B_z)$ and $\hat{n}^j$ is the associated  unit vector. By
a perturbation expansion
in $\epsilon_{k}=\delta B_k^j/B_z$, $k=x,y$, we obtain
\begin{eqnarray}
    U_j &=& \cos(B_z\tau) - i \sigma_z \sin(B_z\tau) \nonumber\\
    & - & i (\sigma_x \frac{\delta B_x^j}{B_z}
    + \sigma_y \frac{\delta B_y^j}{B_z}) \sin(B_z\tau) +
    O\left(\epsilon_{k}^{2} \right) \nonumber
\end{eqnarray}
Note that the zero order terms are nothing but $U^{0}(\tau)$. We shall
denote by  $\delta U_j$ the first order terms, i.e.,
$$\delta U_j\equiv  i (\sigma_x \frac{\delta B_x^j}{B_z}
    + \sigma_y \frac{\delta B_y^j}{B_z}) \sin(B_z\tau) $$

Under the assumption that
$\epsilon_{k}=\delta B_k^j/B_z \ll 1$ the operator $U(T)$ can be easily
computed by first order perturbation theory: by keeping in (\ref{cron})
only the terms that are first order in $\epsilon_{k}$ we obtain
\begin{equation}
    U(T) = U^{0}(T) - \sum_j \delta P^j
\label{eq:evolutor_general_noise}
\end{equation}
where $U^{0}(T)= \exp(-i TH_{0})$ and
$$\delta P^j = \left(U^{0}(\tau)\right)^{j-1} \delta U_j
\left(U^{0}(\tau)\right)^{N-j}$$

To compute the fidelity (\ref{eq:fidelity_def}) for the (generic)
initial state $|\psi(0)\rangle = \alpha |\uparrow\rangle
+ \beta |\downarrow\rangle$ we need to calculate the scalar product
$ \langle \psi^{(0)}(T) |\psi (T)\rangle =\langle U^{0}(T)\psi(0)|
U(T)\psi (0)\rangle$. According to
(\ref{eq:evolutor_general_noise}), this is given by  $$  1- \langle
\psi (0)| (U^{0}(T))^{\dagger}
(\sum_j \delta P^j)|\psi (0) \rangle\, $$
The matrix elements of $\delta P^j$ in the logical basis are:
$(\sum_j \delta P^j)_{mm} = 0$ and
$$(\sum_j \delta P^j)_{lm} = \frac{\sin(B_z\tau)}{B_z}
\sum_j e^{i B_z\tau (N-2j+1)}(\delta B_x^j \pm i \delta B_y^j)\, . $$
By taking into account that $\sin(B_z\tau) \approx B_z\tau$ and
$\exp(i B_z \tau) \approx 1$ for $\tau =
T/N\ll 1$, we may finally evaluate the modulus of the scalar product
and compute the fidelity. We obtain
\begin{equation}
    \mathcal{F}_{T} = \left|1- 2 T \mbox{Re} \left(  \sum_j e^{i 2 B_z (T-\tau
j)}
    (\frac{\delta B_x^j}{N} - i \frac{\delta B_y^j}{N}) \alpha^*
\beta\right) \right|
    \label{eq:fid_non_diagonal}
\end{equation}

It is important to note that, also in this case,
for $N \rightarrow \infty$ and $\tau\to 0$ (while keeping $N\tau =T$ of
order 1),  the  fidelity approaches $1$. This is so because $\delta
B_k^j$ are  independent random variables  with mean zero \cite{prove}.

So, also in this case there is a
noise cancellation effect. This effect is confirmed  by the numerical
simulations shown in Figure \ref{fig:non_diagonal}.

\section{Adiabatic evolution}
\label{sec:adiabatic}

\begin{figure}[t]
    \begin{center}
       \includegraphics[height=5cm]{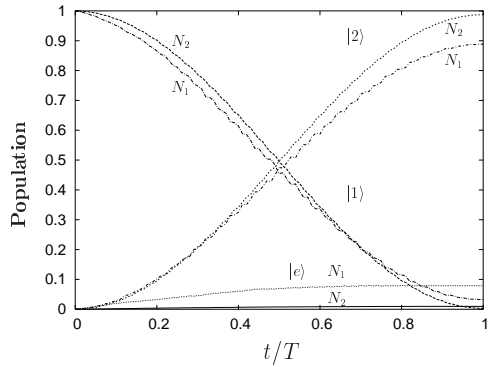}
      \caption{\label{fig:adiabatic_pop} Figure shows the population
        evolution of the three states during the adiabatic transfer
	process for two different correlation time noise ($\tau_i = T/ N_i$).
	For the first curve $N_1= 10^4$ and for the second one $N_2= 10^5$
	for $T=100$ (in arbitrary units).}
    \end{center}
\end{figure}

The {\it adiabatic population transfer} is an important technique used in atoms
optics to achieve population transfer between quantum states of atoms
and molecules \cite{bergman}.
We first create coherence between the initial and
final state ({\it population trapping}) and then produce an
adiabatic evolution to transfer the population to the final state.
This scheme has seen a great success and has been used in many different
areas: chemical reaction \cite{dittmann}, laser-induced cooling
\cite{kulin}, atoms optics \cite{weitz}, cavity quantum electrodynamics
\cite{parkins, walser}.
The wide range of application is due to many advantages of this
scheme: it is easy to implement in different system, it has an high
rate of population transfer and it is robust respect to variations of
field parameters \cite{yatsenko}.

Consider two states $|1\rangle$ and $|2\rangle$
coupled to an excited state $|e\rangle$ by two lasers (i.e., a $\Lambda$
system). The states $|1\rangle$ and $|2\rangle$ can be degenerate or 
quasi-degenerate but it is important that we can address separately both of 
them.
The Hamiltonian in the rotating frame with resonant laser
frequencies is

\begin{equation}
    H = -(\Omega_1(t) |1\rangle \langle e| + \Omega_2(t) |2\rangle
    \langle e|) + h.c. 
    \label{eq:lambda_ham}
\end{equation}
where $\Omega_1(t)$ and $\Omega_2(t)$ are the time-dependent
Rabi frequencies and depend on the parameters of the lasers
(amplitude and phase).
The diagonalization of (\ref{eq:lambda_ham}) gives two eigenstates
$|B_{\pm} \rangle = \frac{1}{\sqrt{2} \Omega}
(\pm \Omega |e\rangle + \Omega_1 |1 \rangle + \Omega_2 |2 \rangle$
(called {\it bright states}) respectively with eigenvalues
$\pm \Omega(t)= \pm \sqrt{\sum_{i=1}^2|\Omega_i|^2}$,
and an eigenstate $|D\rangle = 1/\Omega( \Omega_2
|1\rangle - \Omega_1 |2\rangle )$ (called {\it dark state})
with zero eigenvalue.
In the adiabatic evolution (i.e. when the $\Omega_i$'s change slowly and
$\Omega T \gg 1$) it follows from the adiabatic theorem \cite{messiah}
that if the system starts at time $t=0$ in an
eigenstate of $H(0)$ ({\it dark} or {\it bright} state) during all the
evolution it will remain in the eigenstate of $H(t)$ with the same
eigenvalue.

Now we provide a simple example of the foregoing.
Suppose that $\Omega_1(t)$ and $\Omega_2(t)$ are such that $\Omega$
is time-independent and $\Omega_1(0) =0$ and $\Omega_2(0)= \Omega$.
Moreover, suppose that the initial state is $|1\rangle$.
Then {\it slowly} turn on the first Rabi frequency and turn off the
second one.
The system will always be in the {\it dark state} $|D \rangle$ and,
at the end of the evolution (i.e., when $\Omega_1(T) = \Omega$ and
$\Omega_2(T)=0$), we will be in $|2\rangle$ state and have achieved
population
transfer from $|1\rangle$ to $|2\rangle$.

\begin{figure}[t]
  \begin{center}    
    \includegraphics[height=5cm]{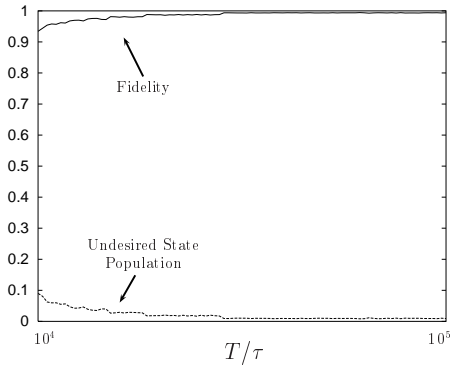}
      \caption{\label{fig:adiabatic_fid}
      Figure shows the {\it fidelity} and the population of exited state
      $|e\rangle$ as function of $T/\tau = N$.}
    \end{center}
\end{figure}

Consider now the case in which we have not a complete control of the laser
field but the Rabi frequencies can fluctuate
$\Omega_i \rightarrow \Omega_i +\delta \Omega_i$ (where the $\delta
\Omega_i$ are independent and Gaussian distributed with zero mean and variance
$\sigma^2$).
This effect can produce errors in population transfer scheme for two
reason: in general noisy perturbations may yield to significantly different
output state and fast fluctuations could break the adiabatic approximation
leading to transition to undesired ({\it bright}) states.

Numerical simulations show that, again,
in the fast fluctuation regime the noise effects cancel out.
In Figure \ref{fig:adiabatic_pop} we present the population evolution of
$|1 \rangle$, $|2 \rangle$ and $|e \rangle$ states
subject to noisy evolution during the
$|1 \rangle \rightarrow |2 \rangle$ {\it coherent adiabatic transfer}
\cite{adiabatic_sims}.
More precisely, we start from $|1\rangle$ and, during the evolution, the
$|2\rangle$ is populated; at the end only the $|2\rangle$ is present.
The $|e \rangle$ is never populated because of the high value of
$\Omega T$ parameter.
These simulations are done using $\Omega T = 1000 $.
Since we are in the adiabatic regime we are sure that the errors
present are those induced by the perturbations.
In Figure \ref{fig:adiabatic_pop} we show the population evolution
when the system is subject to noise with different correlation time $\tau$ (and
different $N = T/\tau$).
For $N = 10^4$ the transfer operation is not precise (i.e. the
$|2\rangle$ state is not completely populated) and the $|e \rangle$ state is
populated; this is a sign of the breaking of the adiabatic approximation due to
noise.
For $N = 10^5$ the evolution is much more similar to the ideal one:
$|2\rangle$ is completely populated and $|e \rangle$ never appears
during the evolution.

The above results can be explained calculating the amplitude transition
from the {\it dark} to the {\it bright} state in presence of  
perturbations.
The standard rule (see, e.g. \cite{schiff}) to calculate
the probability amplitude of a transition from the $n$ state to one of the
$k$ state ($k \neq n$) at time $t$ is 
$a_k = \sum_n^{\prime} \int_{-\infty}^t \frac{a_n(t)}{\omega_{kn}}
\frac{\partial V_{kn}}{\partial t}
exp({i \int_0^t \omega_{kn} t^{\prime}}) dt^{\prime} $
where $\sum^{\prime}$  indicates that the term $n=k$ is omitted,
$\omega_{kn} = E_k -E_n$ and $V_{kn}$ is the matrix element associate to
the transition $n \rightarrow k$.
In our case, the initial state is the {\it dark state} and the final
states are the {\it bright states}. The eigenvalues $\pm \Omega$ are
constant and then $\omega_{kn} = \Omega$.
The perturbation Hamiltonian for time $j\tau \leq t \leq (j+1) \tau$
in the $|e\rangle$, $|1\rangle$, $|2\rangle$
basis is $\delta V^j = \sum_{i=1}^2 \delta \Omega_i^j
|i\rangle \langle e| + h.c.$
and in the new ({\it dark-bright states}) basis the relevant matrix element
are

\begin{equation}
V^j_{DB_+} = - V^j_{DB_-} =\sum_j \frac{\delta \Omega_2^j \Omega_1 -
\delta \Omega_1^j \Omega_2 }{\sqrt{2} \Omega} S(j \tau , (j+1) \tau)
\label{eq:pert_ham}
\end{equation}

To calculate the matrix element $\frac{\partial V_{kn}}{\partial t}$
we must take into account that $\partial S(j \tau , (j+1) \tau)/
\partial t = \delta (t- j \tau) - \delta (t- (j+1) \tau)$.
Let us focus our attention only on the first term in (\ref{eq:pert_ham}),
by differentiating it we obtain
$1/(\sqrt{2} \Omega) \sum_j \delta \Omega_2^j (\partial
\Omega_1/\partial t)
S(j \tau , (j+1) \tau) - \Omega_1 \delta \Omega_2^j
(\delta(t-j \tau) -\delta(t-(j+1) \tau)$.
The terms $\partial \Omega_i/\partial t$ representing the adiabatic  
driven evolution are very small and can be neglected.
Inserting this result in the expression for $a_k$ and performing the
integration we have

\begin{eqnarray}
    a_k & = & \frac{1}{\sqrt{2} \Omega^2} \sum_j \delta \Omega_2^j
    [ a_n(j \tau) \Omega_1 (j \tau) e^{i\Omega j \tau} \nonumber \\
      & - &  a_n((j+1) \tau) \Omega_1 ((j+1) \tau)
      e^{i \Omega (j+1) \tau} +  \nonumber \\
      & + & \mbox{terms with } (\Omega_1 \leftrightarrow  \Omega_2 )
      \label{eq:trans_amp}
\end{eqnarray}
If $a_n$ and $\Omega_i$ change slowly i.e.,
$a_n((j+1)\tau) \approx a_n(j \tau)$
and $\Omega_i ((j+1) \tau) \approx \Omega_1 (j \tau)$.
The exponential terms can be simplified to obtain
$\exp(i \Omega \tau (j+ 1/2)) \sin(\Omega \tau/2)/(2i)$ and  eq.
(\ref{eq:trans_amp}) for $\tau = T/N \ll 1$ gives

\begin{equation}
    a_k = \frac{T}{2\sqrt{2} i \Omega} \sum_j e^{i\Omega j \tau} a_n(j
\tau)
    \left[ \Omega_1 (j \tau)  \frac{\delta \Omega_2^j}{N} - \Omega_2 (j
\tau)
    \frac{\delta \Omega_1^j}{N}\right]
\end{equation}

The sums of $\frac{\delta \Omega_i^j}{N}$ converge to the mean of
$\delta \Omega_i$ that is to zero;
the other factors are bounded ($0 \leq |a_n| \leq 1$
and $0 \leq |\Omega_i| \leq \Omega $) and by consideration similar to
those at the end of section \ref{sec:non_diagonal} (and \cite{prove})
we can conclude that $a_k \rightarrow 0$ for $N \rightarrow \infty$.
In our case, $|n\rangle = |D\rangle$,
$|k \rangle = |B_{\pm} \rangle$ and $|a_{B_{\pm}}|^2 \rightarrow 0$ as
$N \rightarrow \infty$ : the transition from {\it dark} to {\it bright
states} is suppressed in the fast fluctuating regime
and the evolution happen in the {\it dark space}.

To have a more detailed picture in Figure \ref{fig:adiabatic_fid}
we show also the trend of the {\it fidelity}
(upper curve) and the relative average population of the $|e \rangle$
state
(lower curve) as functions of $T/\tau$.
The trends of the two curves are correlated, which is suggesting that
the main source of error in the operation is the population of excited state 
due to loss of the adiabatic approximation.
As expected, because of the cancellation effect, for great $T/\tau$ the 
{\it fidelity} approaches $1$ even in presence of strong noise and
the $|e \rangle$ state is not populated.

These results not only can explain why the {\it adiabatic population
transfer}
scheme is robust against field fluctuation but can give information for
the
experimental set-up.
In fact the adiabatic evolution is, in general, arbitrarily long;
once the experimental parameters are fixed (as the laser with its proper
noise correlation time $\tau$) we can prolong the evolution time in
order
to increase $T/ \tau$ and let the noise average out.
As shown before, for every noise strength $\sigma$
and correlation time $\tau$ we can find and evolution time $T$
in order to obtain the desired {\it fidelity}:
that is, to achieve the population transfer with arbitrary small error.

\section{Conclusion}
\label{sec:conclusion}

We studied the effect of stochastic noise on several quantum systems.
The noise is described by a Gaussian stochastic process with zero mean
superposed to the ideal quantum evolution.
For all of the systems we found that for fast fluctuating noise a
{\it cancellation effect} appears: the noise fluctuations average out
leading the system to a state near to the ideal one.

We showed by analytical and numerical calculation how this effect can appear in
a two level system and propose an experiment to verify the presence of
this cancellation regime. The experiment is based on the one performed
in Ref. \cite{kielpinski} and we think that, with a modification of the
experimental set-up, it could be easily performed.

We applied the same model to another important technique in atoms
physics : the {\it adiabatic population transfer}.
We explained how this effect leads to the robustness of the
adiabatic process against the perturbation noise.
This can be important for the experiments using the
{\it adiabatic population transfer}.

\begin{acknowledgments}
P.S. wishes to thank D. Kielpinski for useful comments.
\end{acknowledgments}


\begin{thebibliography}{99}
\bibitem{sample}
Our sampling set of the Bloch sphere (i.e. our inizial state space)
is given by
$\{ \pm e_i\}_{i=x,y,z}, (\pm e_x\pm e_y)/\sqrt{2},
(\pm e_x\pm e_z)/\sqrt{2}, (\pm e_y\pm e_z)/\sqrt{2}$.
Here $e_i$ denotes the normalized vector of the $i$-th direction.
\bibitem{note_on_realizations} We make simulations with $1000$
different realizations and average the resulting {\it square fidelity}.
A part from a noisier in the curve in the plots,
the results are weakly affected by this parameters so that with $100$
realizations we obtain the almost the same results.
\bibitem{prove}
  This is an elementary instance of the law of large numbers:
  Suppose to have a sum written in the form
  $\sum_{j=0}^N f(\tau j) \frac{x_j}{N}$ where $x_j$ are independent
  random variable with zero mean and variance $\sigma^2$,
  $\tau=T/N$ and $0 \leq j \leq T$.
  $Var(Y_N)$ (the variance of $Y_N= \frac{1}{N} \sum_{j=0}^N
  f(\tau j)x_j$) is $\frac{\sigma^2}{N^2} \sum_{j=0}^N (f(\tau j))^2$.
  Thus, if $f$ is bounded in the interval $[0,T]$, $Var(Y_N) \rightarrow 0$
  when $N \rightarrow \infty$ and therefore $Y_N \rightarrow 0$
  since the $x_j$ mean is zero.
  Since in our case $f(\tau j) = exp(-i 2 B_z \tau j)$ the second terms in
  eq. (\ref{eq:fid_non_diagonal}) goes to zero and $\mathcal{F} \rightarrow 1$
  for $N \rightarrow \infty$ for every initial state
  (i.e. independently from $\alpha$ and $\beta$).
\end{thebibliography}
\end{document}